\documentclass[a4paper,11pt]{article}
\usepackage{jinstpub}
\notoc

\title{The optimization, design and performance of the FBCM23 ASIC for the upgraded CMS beam monitoring system}

\author[a]     {Jan~Kaplon}
\author[a,1] {, Grzegorz~Wegrzyn\note{Corresponding author.}}
\author[a,b]     {, Konstantin~Shibin}
\author[a,c] {, Marnix Barendregt}

\affiliation[a]{
	CERN,\\
	Geneva, Switzerland}

\affiliation[b]{
	TALTECH\\
	Tallinn, Estonia}

\affiliation[c]{
    University of Cambridge\\
    Cambridge, UK
}

\emailAdd{grzegorz.jan.wegrzyn@cern.ch}

\abstract{
We present the development of the FBCM23 ASIC designed for the Phase-II upgrade of the Fast Beam Condition Monitoring (FBCM) system built at the CMS experiment which will replace the present luminometer based on the BCM1F ASIC \cite{Przyborowski:2016aco}. 
The FBCM system should provide reliable luminosity measurement with 1\,ns time resolution enabling the detection of beam-induced background. 
The FBCM23 ASIC comprises 6 channels of the fast front-end amplifier working in transimpedance configuration, booster amplifier, and leading edge discriminator. The complete processing chain provides an overall shaping function equivalent to the CR-RC$^3$ filter.
The paper will show the optimization of the design, overall architecture, and the detailed implementation in a CMOS 65\,nm process as well as preliminary electrical performance.
}


\begin{document}
\maketitle

\section{Specifications and  optimisation of the design\label{Requirements} }
\begin{figure}[htbp]
\begin{minipage}[t]{0.45\linewidth}
    \includegraphics[width=\linewidth]{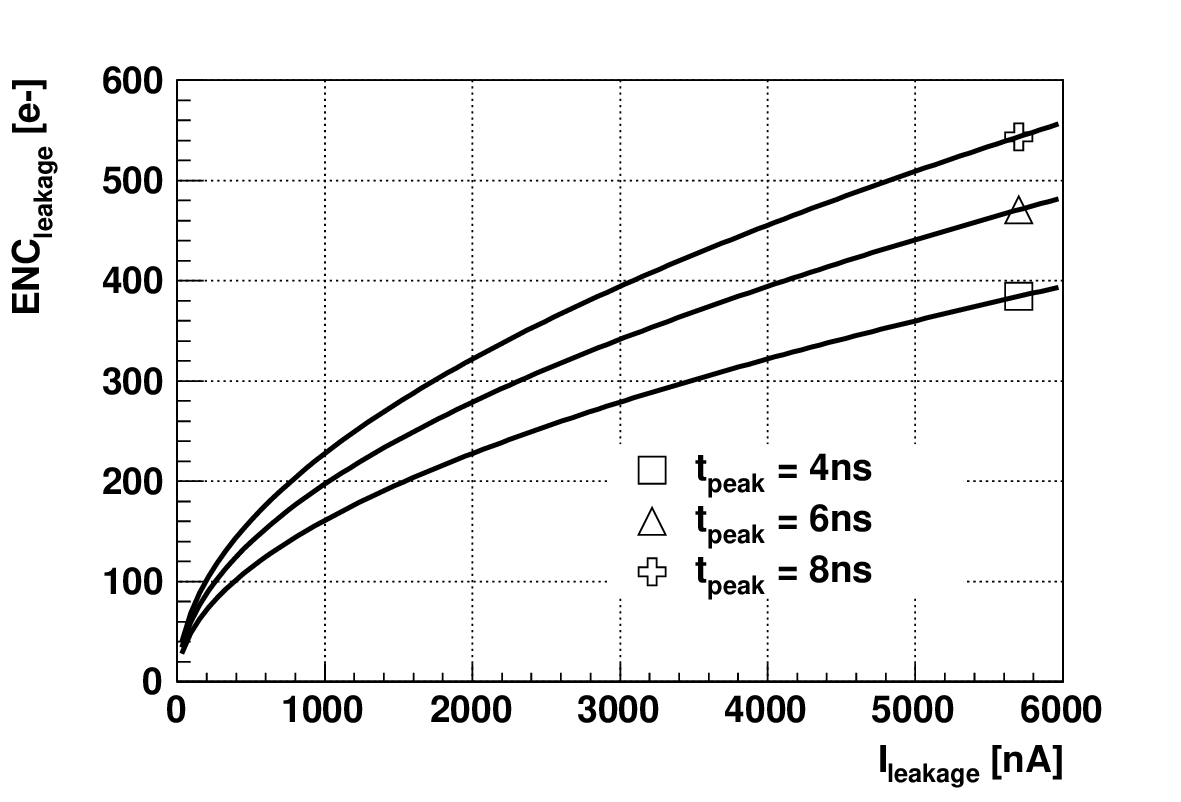}
    \caption{Noise contribution from the sensor leakage current for 6 and 8\,ns peaking time and the CR-RC$^3$ shaping.}
    \label{fig:ENCsensorLeak}
\end{minipage}%
    \hfill%
    \begin{minipage}[t]{0.45\linewidth}
    \includegraphics[width=\linewidth]{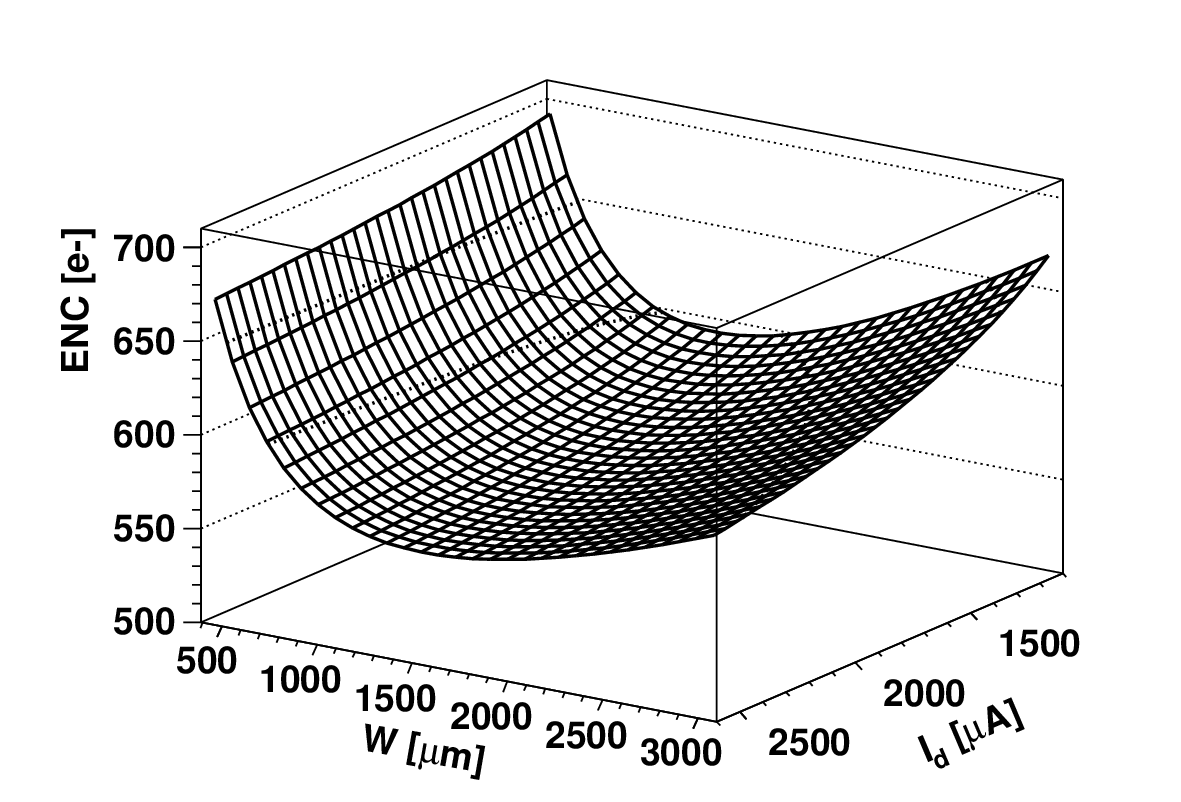}
    \caption{Series noise contribution for 8\,ns peaking time and the CR-RC$^3$ shaping for 5\,pF input capacitance (-20$^{\circ}$C).}
    \label{fig:ENCseries}
\end{minipage}%
\end{figure}

\begin{figure}[htbp]
\begin{minipage}[t]{0.45\linewidth}
    \includegraphics[width=\linewidth]{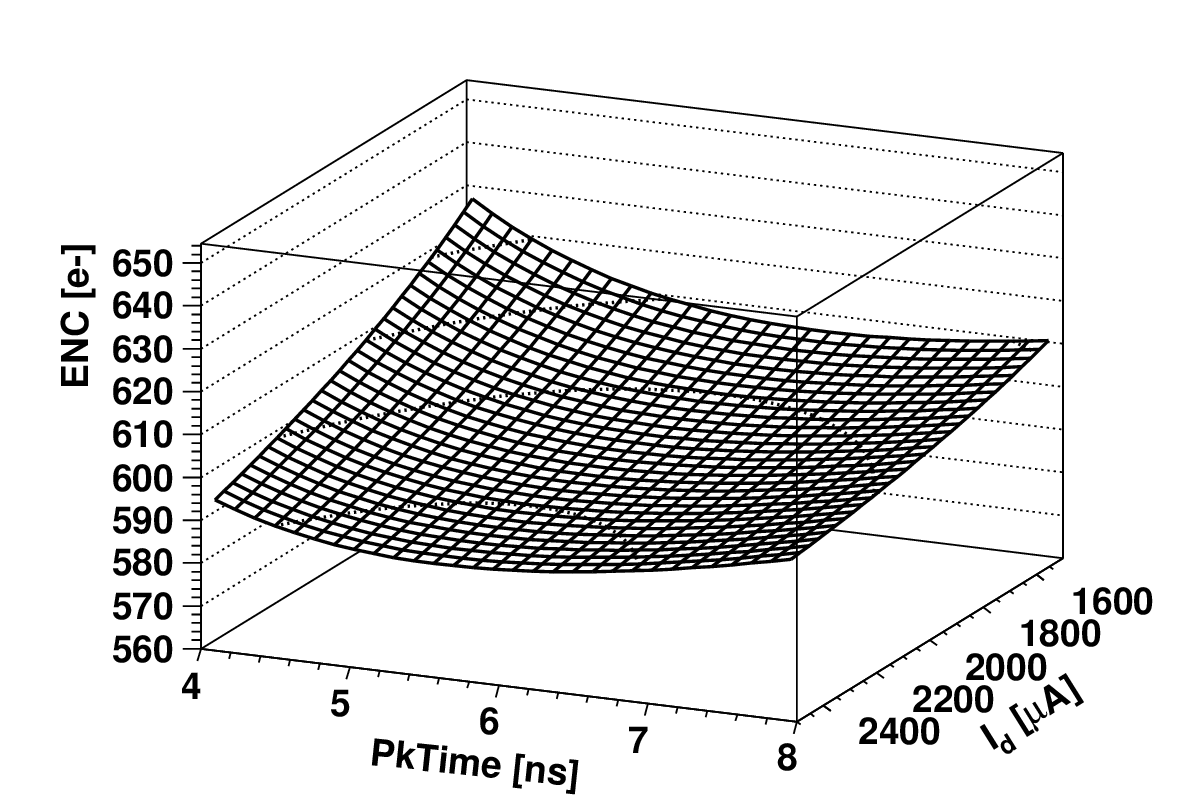}
    \caption{The total ENC figure showing optimum peaking time assuming 3\,pF input capacitance and 2\,$\mu$A leakage current at -20$^{\circ}$C operation.}
    \label{fig:ENC3D3p2uA}
\end{minipage} 
    \hfill%
\begin{minipage}[t]{0.45\linewidth}
    \includegraphics[width=\linewidth]{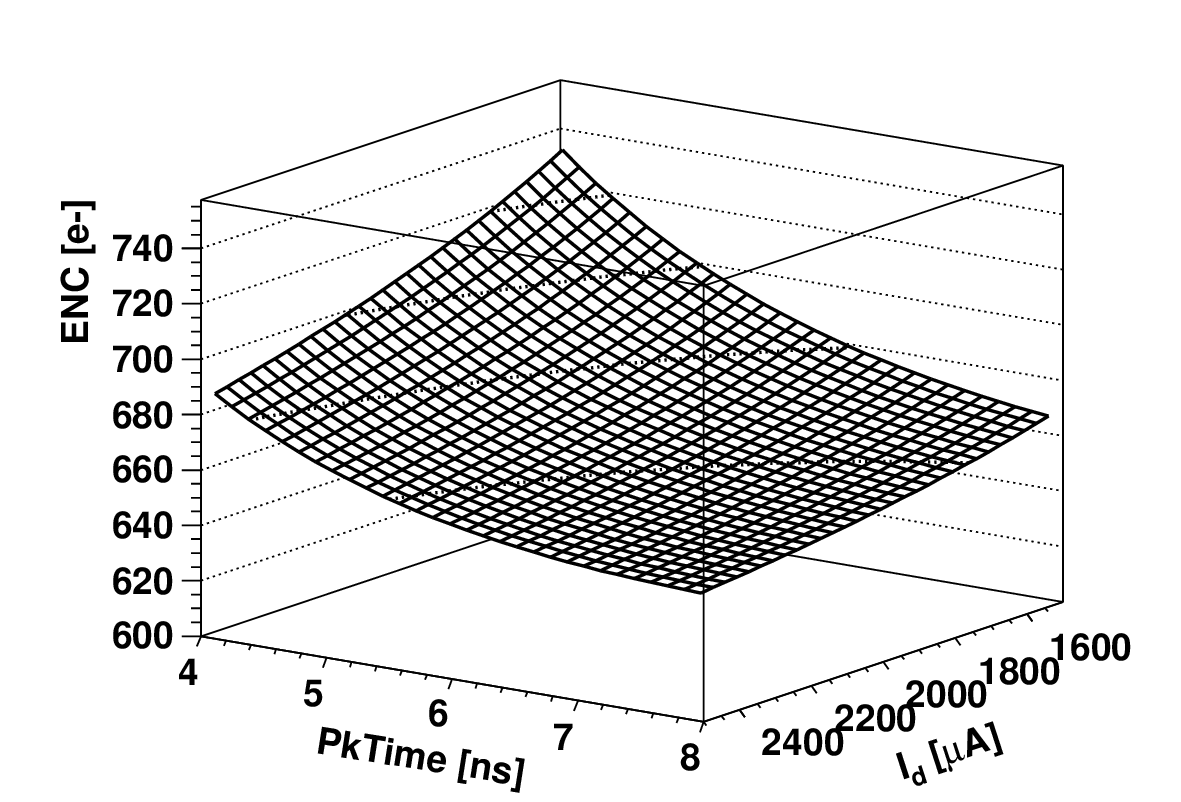}
    \caption{The total ENC figure for 5\,pF input capacitance and 1\,$\mu$A leakage current (-20$^{\circ}$C).}
    \label{fig:ENC3D5pF1uA}
\end{minipage} 
\end{figure}

The primary role of the FBCM \cite{Sedghi_2022} system is accurate luminosity measurement with 1\,ns time resolution that will provide the detection of beam-induced background. 
To meet this requirement from the physics point of view, the FBCM sensors should have a particular area and distance to the beamline, balancing occupancy, and acceptance. 
A good compromise between the area and the position of the sensor in the experiment is provided by a 1.7$\times$1.7\,mm silicon pad installed at a radius of around 14.5\,cm \cite{Collaboration:2759074}.  
The radiation environment in this position is rather harsh, and the detector modules should stand up to 200\,Mrad of total ionizing dose (TID) and particle fluxes up to 2.5$\times$10$^{15}$\,N/cm$^2$\,1\,MeV equivalent. 
Although the 65 nm CMOS process can stand the expected TID dose without major issues, the increase of the leakage and degradation of charge collection efficiency (CCE) from the heavily irradiated sensors have to be taken into account during the noise optimization of the front-end amplifier and final choice of the sensor thickness.\\ 
For the FBCM detector, a cost-effective solution will be to produce the sensors using either 290\,$\mu$m material of the outer tracker or 150\,$\mu$m wafers used by the pixel detector. Both solutions have advantages and drawbacks, and the FBCM23 ASIC has to support both options.
The expected numbers for the leakage currents are in the micro-Amper range. 
The worst case values should be estimated using the figures published in \cite{LINDSTROM200330} or \cite{WONSAK2015126} and scaled to the intended operating temperature of -35$^{\circ}$C \footnote{The figures presented in the literature are normalized to +20$^{\circ}$C. The typical setups used for the characterization of the irradiated sensors work down to -20$^{\circ}$C}. 
Operating the sensors at such temperatures sets some restrictions on the power dissipated by the ASIC. Although there are no hard constraints from the point of view of the designed cooling and power supply systems, it will be reasonable to limit the power dissipation of the ASIC below 100\,mW and, consequently minimize expected temperature gradients on the detector module.
For the 150\,$\mu$m sensor working at -20$^{\circ}$C after  2.5$\times$10$^{15}$\,N/cm$^2$ one can expect the leakage in the range of 2\,$\mu$A. 
The same leakage will occur for the 290\,$\mu$m sensor and fluence of 1$\times$10$^{15}$\,N/cm$^2$, which is most likely the maximum allowable dose from the point of view of CCE dropping to about 50\,\% (see predictions in \cite{Akchurin_2020}). 
That means that FBCM detector built with 290\,$\mu$m sensors has to be replaced in the middle of the operation, which in fact will be possible on the occasion of the replacement of the innermost layer of the pixel detector. 
Although the degradation of the CCE and leakage will be lower for the thinner sensor (0.95 and 1\,$\mu$A respectively after 1$\times$10$^{15}$\,N/cm$^2$, see \cite{Akchurin_2020}), one should keep in mind that it will also provide a smaller charge before irradiation as well as it will show higher capacitance (around 4\,pF to be compared to 2\,pF for 290\,$\mu$m sensor).
The contribution of the shot noise coming from the sensor leakage to the total ENC is shown in Fig. \ref{fig:ENCsensorLeak}. 
One can see that for the shaping time in the range of 4 to 8\,ns and for the expected leakages below 3\,$\mu$A, this contribution can be kept at a reasonable level, under 400\,e$^-$\,ENC.\\ 
One can stress that the fast shaping times are important not only from the standpoint of the optimal filtering of the noise contributed by the sensor leakage but also from the point of view of the timing requirements. 
The time-walk for the leading edge discriminator (LED), which is the most robust solution for the binary system, connected to 8\,ns peaking time CR-RC$^3$ shaper will be below 5\,ns. 
This is what is compatible with the requirements of 1\,ns rms time resolution. \\
In order to provide a reasonable Signal-to-Noise Ratio (SNR) above 10 at the end of the detector lifetime for any sensor option, the series noise contribution from the input transistor should be kept below 700\,e$^-$\,ENC. Figure \ref{fig:ENCseries}  shows the optimization\footnote{All noise calculations were performed using models described in \cite{1589268} and \cite{6220879}.} of the input transistor dimensions and the bias for the worst case of 5\,pF input capacitance (4\,pF sensor + 1\,pF contingency for the parasitics).
One can see that for 2000\,$\mu$m width of the input transistor, chosen for the FBCM23 input stage, biased with 2\,mA current, the maximum contribution of the series noise is below 600\,e$^-$\,ENC.
Figure \ref{fig:ENC3D3p2uA} and \ref{fig:ENC3D5pF1uA} show the optimum peaking times for both options of the sensor thickness irradiated up to 1$\times$10$^{15}$\,N/cm$^2$. 
While the optimal peaking time for the 2\,$\mu$A leakage is around 5\,ns from the ENC standpoint (see Figure~\ref{fig:ENC3D3p2uA} ), one should also watch the possible degradation of CCE due to ballistic deficit effect in the heavily irradiated sensor. 
For thinner sensors and lower leakage currents, the optimal peaking time is closer to 8\,ns. Its further increase will be limited by the requested time resolution of the ASIC.

\section{ASIC implementation \label{architecture} }
The binary architecture employed in the FBCM23 ASIC provides simplification of the off-detector electronics and allows for direct interfacing to the LPGBT chip (\cite{Moreira:2809058}), which is the core of the CERN standard digital transmission system. 
Because of the limited number of channels on the detector module (6), it is possible to read each single FBCM23 channel output by the LPGBT input, which is capable of sampling the data with 800\,ps time bin resolution. 
This approach simplifies the design of the FBCM23 ASIC, which consists of six parallel binary front end channels associated with a number of peripheral blocks providing biasing of the amplifier stages, threshold control, and calibration function.
\begin{figure}[htbp]
\begin{minipage}[t]{0.55\linewidth}
    \includegraphics[width=\linewidth]{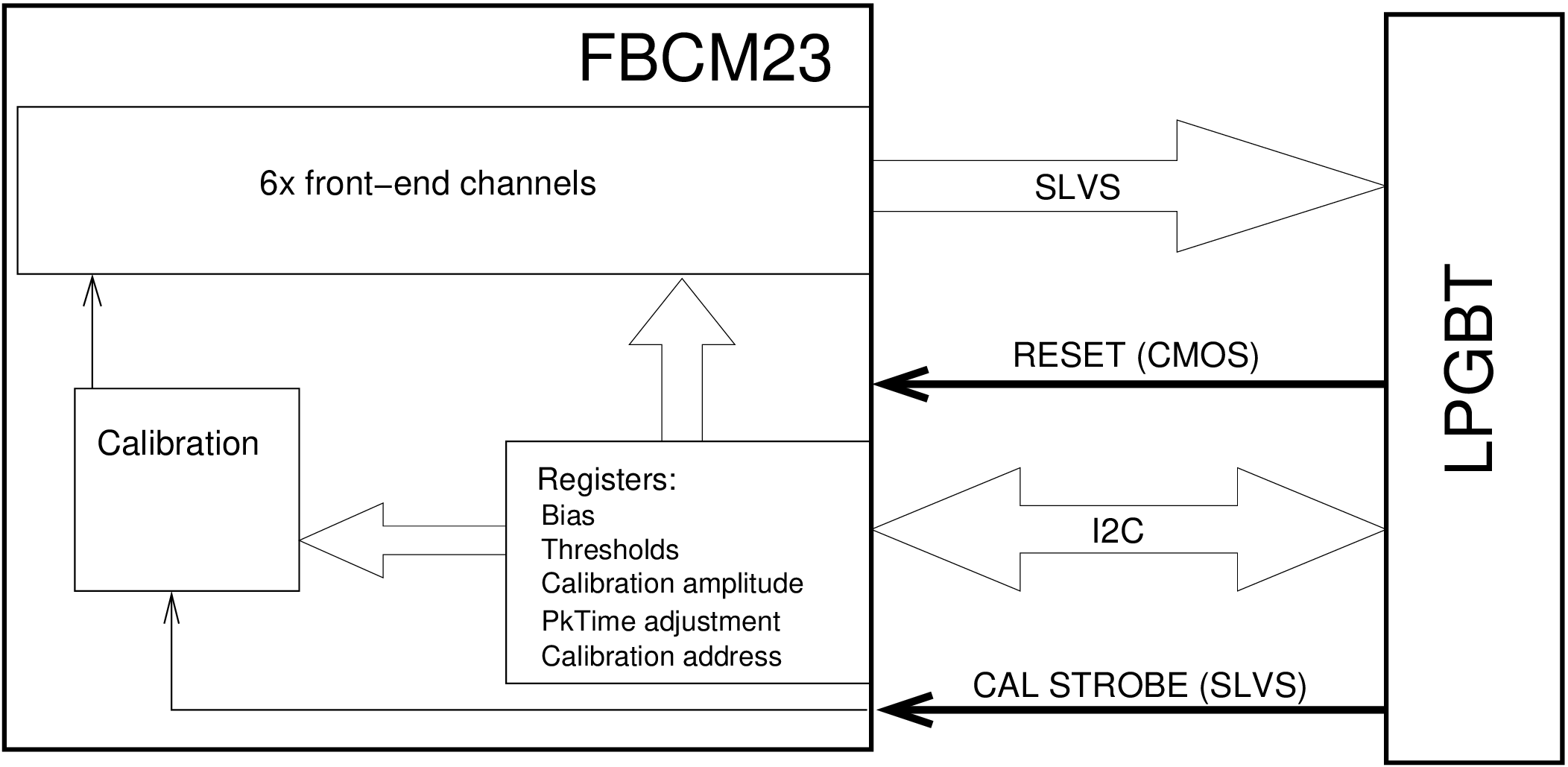}
    \caption{Block diagram of the FBCM23 ASIC.}
    \label{fig:FBCMblock}
\end{minipage}%
    \hfill%
\begin{minipage}[t]{0.34\linewidth}
    \includegraphics[width=\linewidth]{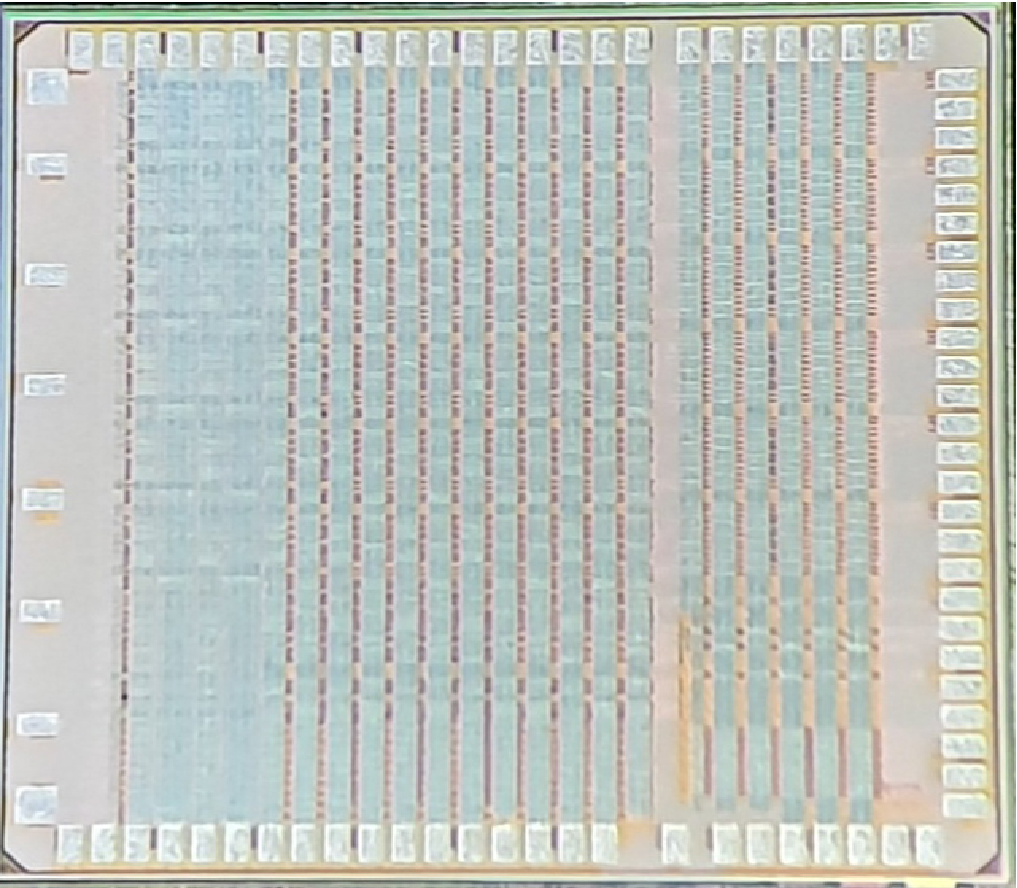}
    \caption{Photo of the FBCM23 ASIC.}
    \label{fig:FBCM23photo}
\end{minipage} 
\end{figure}
The block diagram of the FBCM23 ASIC is shown in Figure \ref{fig:FBCMblock}. The FBCM23 comprises 6 identical channels layouted on the 3$\times3$\,mm$^2$ of silicon area. All bias and configuration registers are accessible via a standard I2C interface.
The chip has been implemented and fabricated in 65\,nm CMOS process.  
During the layouting of the ASIC, considerable attention has been drawn to the correct separation of the analog and digital domains using deep NWELL transistors and guarding techniques reinforced by blocking the substrate doping using the mask intended for the native devices. Special care has been put on the robustness of the power grids implemented with top, thick metals.
On the photograph of the ASIC shown in Figure \ref{fig:FBCM23photo} one can see the top metal vertical power grid occupying about 50\,\% of the top metal area. 
Another thick metal layer is used for the horizontal power distribution along the electronic channels. 
The analog part of FBCM23 consumes for the nominal settings around 32\,mA, from which about 40\,\% is spent in the input transistors.  About 28\,mA is consumed by the digital part, mainly the SLVS output drivers. The simulated DC drops across the power bars are below the 1\,mV range. 
\begin{figure}[htbp]
\begin{minipage}[t]{0.38\linewidth}
    \includegraphics[width=\linewidth]{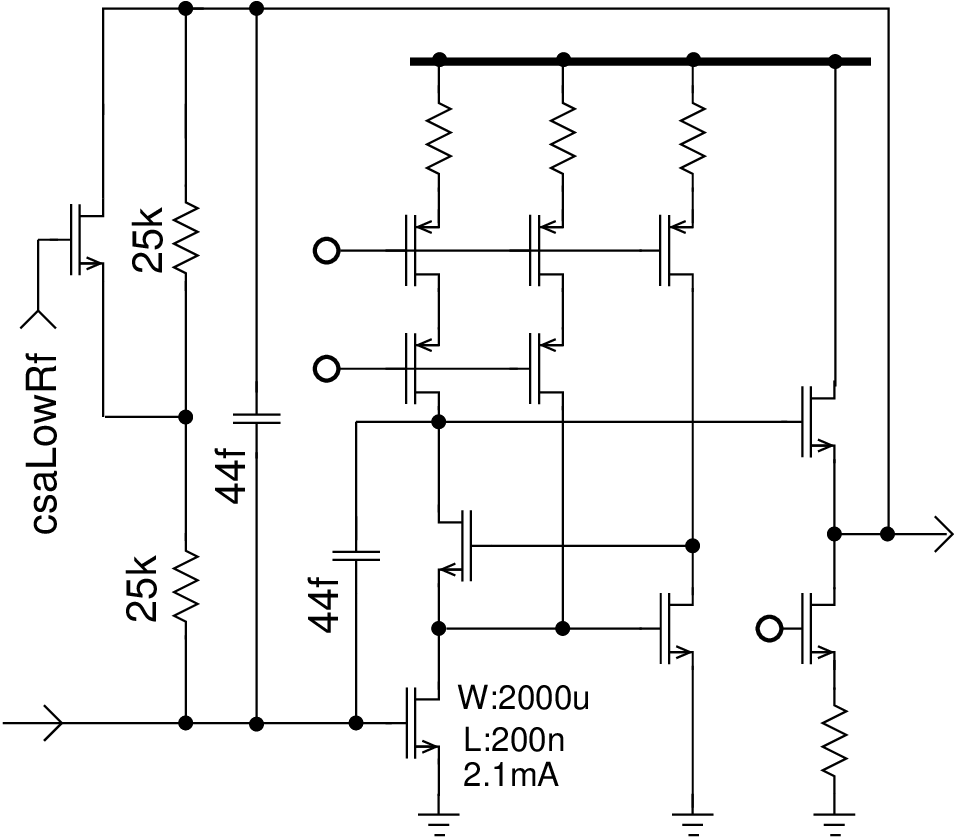}
    \caption{Schematic of the preamplifier.}
    \label{fig:PreampSchem}
\end{minipage} 
\hfill%
\begin{minipage}[t]{0.56\linewidth}
    \includegraphics[width=\linewidth]{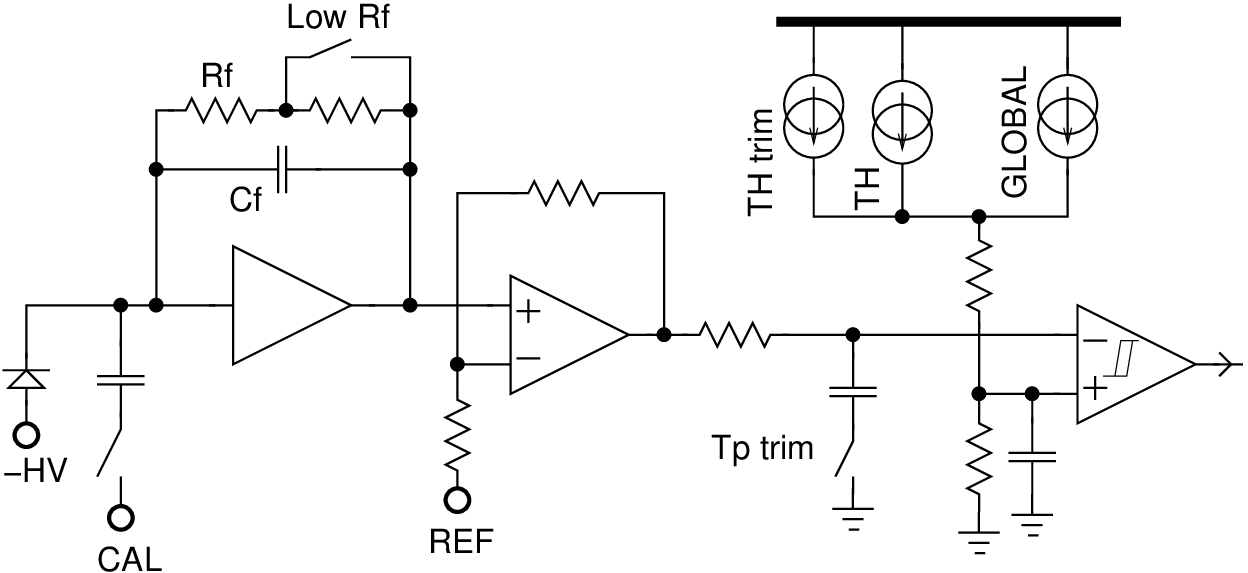}
    \caption{Schematic of one front end channel.}
    \label{fig:FEschem}
\end{minipage}%
\end{figure}
\\
The schematic diagram of the preamplifier is shown in Figure \ref{fig:PreampSchem}. The input stage is built with the regulated telescopic cascode amplifier, with the NMOS input transistor of 2000/0.2\,$\mu$m biased with 2.1\,mA and loaded with low voltage cascode PMOS sources. 
The simulated open loop gain and gain-bandwidth product (GBP) are around 69\,dB and 3.5\,GHz respectively, which allows for the fast shaping of the detector pulses.
The preamplifier works in a transimpedance configuration with an adjustable feedback resistor (50 or 25\,k$\Omega$). \\
The simplified schematic diagram of the front end channel is shown in Figure \ref{fig:FEschem}. The transimpedance preamplifier is DC coupled to the booster amplifier and leading edge discriminator.
Although this intrinsically provides good stability of the baseline in case of high and variable hit rates, it is a less satisfactory solution from the standpoint of the mismatch variation which is amplified by DC-coupled amplifier stages. 
The DC variation at discriminator input is nearly 100mV pk-pk and to compensate for this, two 8-bit threshold DAC’s per channel have been employed. 
Another 8-bit DAC, common for the chip, is used for the global offset setting. 
The gain of the full chain for the nominal settings is around 60\,mV/fC. 
The switchable RC filter (RC set range between 0 and 6) provides adjustment of the peaking time between 5 and 9\,ns range, which helps in the optimization of SNR for various input load conditions (sensor capacitance and leakage current). 
The complete processing chain consisting of the preamplifier, booster amplifier, RC filter, and input stage of the discriminator provides an overall shaping function equivalent to the CR-RC$^3$ filter.
The outputs of the discriminators are sent outside the chip through SLVS interfaces.\\
\section{Test results}
\begin{figure}[htbp]   
\begin{minipage}[t]{0.42\linewidth}
    \includegraphics[width=\linewidth]{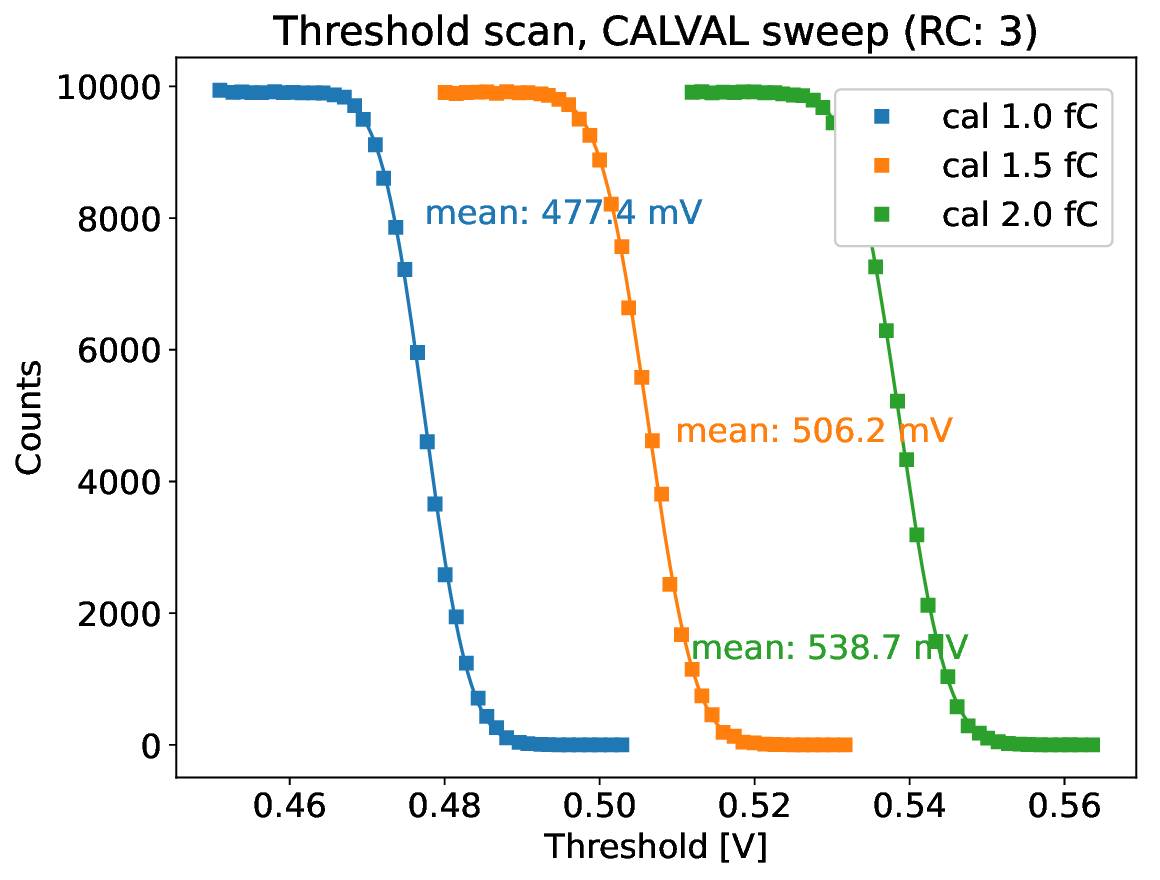}
    \caption{Threshold scans for 1, 1.5, and 2\,fC charge injection and RC set to 3. Channel with unloaded input.}
    \label{fig:thscan}
\end{minipage} 
\hfill%
\begin{minipage}[t]{0.42\linewidth}
    \includegraphics[width=\linewidth]{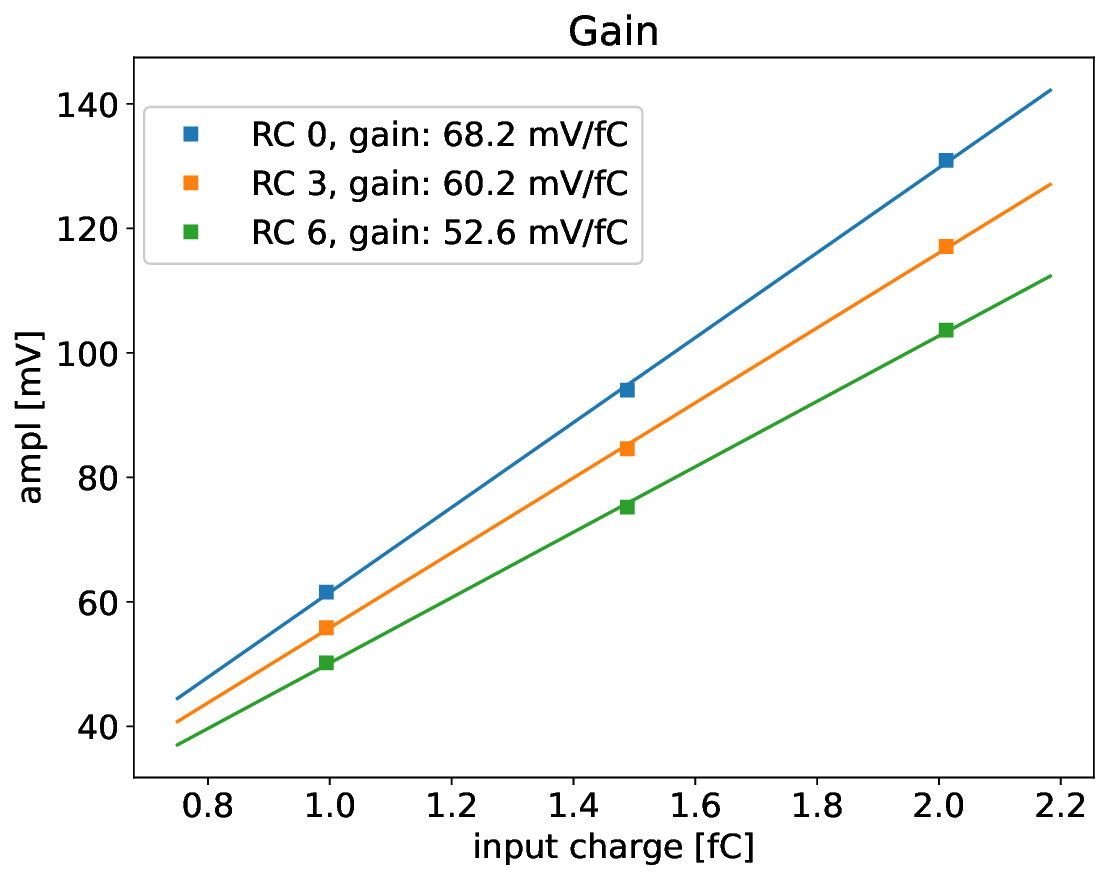}
    \caption{The gain measurement for RC set to 0,3 and 6.}
    \label{fig:gain}
\end{minipage}
\end{figure}

\begin{figure}[htbp]
\begin{minipage}[t]{0.42\linewidth}
    \includegraphics[width=\linewidth]{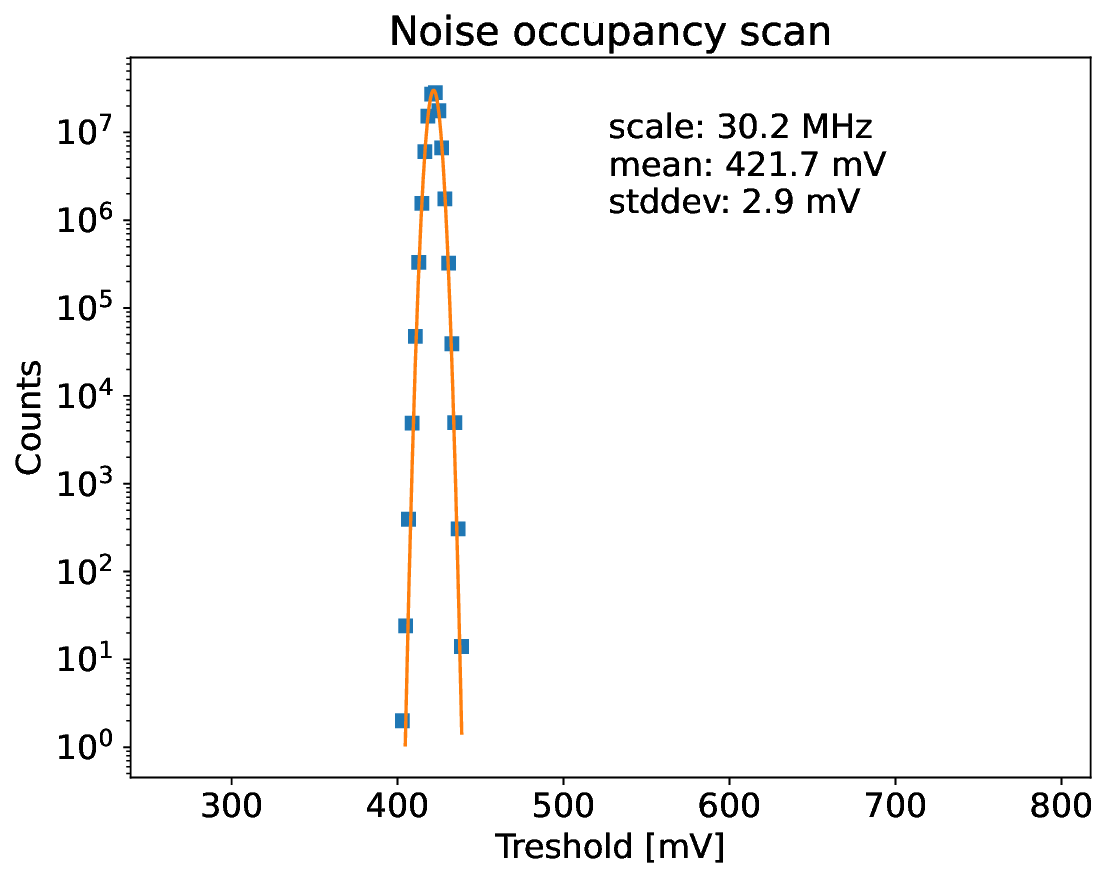}
    \caption{The noise occupancy scan for the channel with unloaded input and RC set to 3 (nominal configuration).}
    \label{fig:noiseOccupancy}
\end{minipage} 
    \hfill%
\begin{minipage}[t]{0.42\linewidth}
    \includegraphics[width=\linewidth]{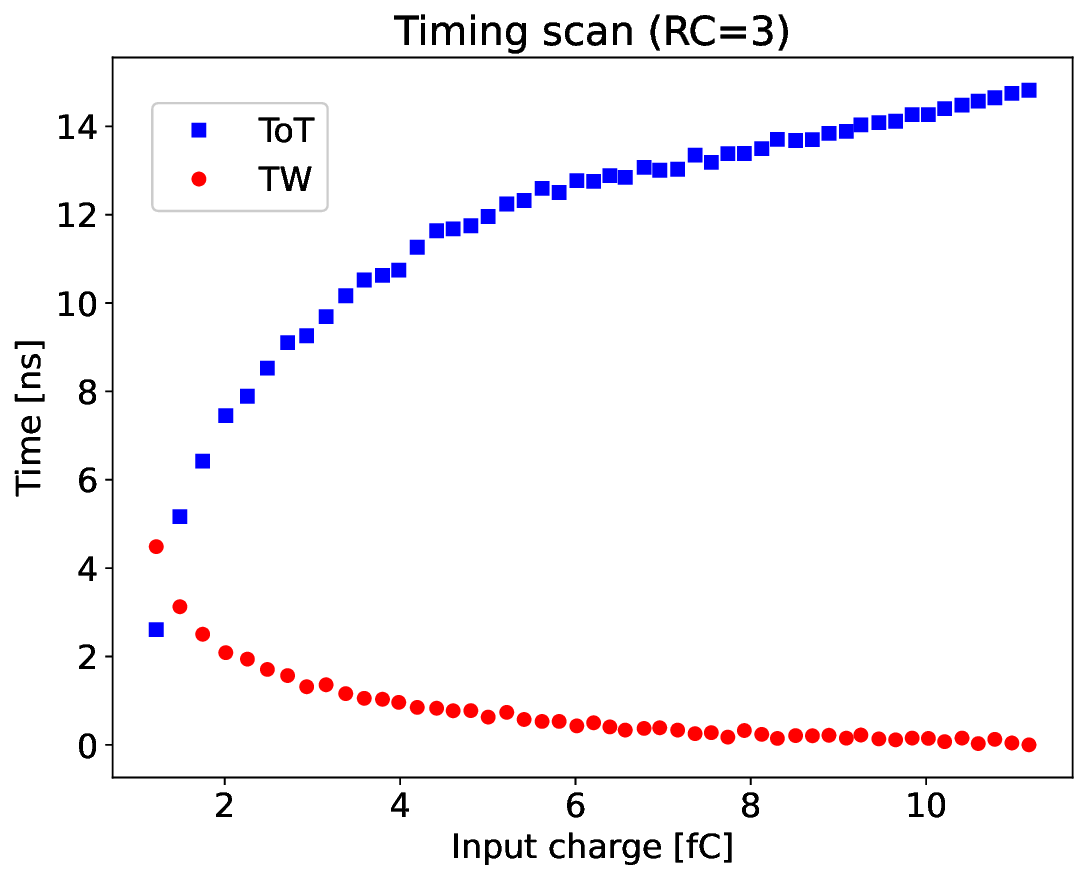}
    \caption{Time walk and time over threshold measurement for RC set to 3 (nominal) and threshold set to 1\,fC.}
    \label{fig:twalk}
\end{minipage} 
\end{figure}

The initial evaluation of the ASIC has been carried out using the internal charge injection circuitry based on an 8-bit DAC supplying a DC current to the resistor which is short-circuited by an NMOS switch controlled by the strobe signal provided by an external pulse generator (Agilent 81110a). 
The calibration charge is injected into the front end amplifier input through on-chip, per channel 52\,fF calibration capacitors.
The output of the single channel is read out by the Agilent 53132A counter (noise and gain measurement) or by the Tektronix scope MSO64B (timing characterization).
Gain and noise are extracted from threshold scan measurements at three different charges,
1, 1.5, and 2\,fC. Each threshold scan is fitted with a complementary error function (S-curve) from which the median and width can be extracted. 
The gain is extracted from the dependence of the median on the injected charge whilst the noise is extracted from the fit to the noise occupancy scan.
The timing variation of the front end as a function of the injected charge (time-walk) is measured as a delay of the output pulse with respect to the strobe signal for different injection charges and the discriminator set to the operating threshold.\\
Figure \ref{fig:thscan} shows the threshold scans performed for 1, 1.5, and 2\,fC injected signals together with the fit to the complementary error function for one particular setting of RC filter. The medians of the fitted S-curves as a function of the injected charge for three different settings of the RC filter are shown in Figure \ref{fig:gain}. The gain extracted from the scans agrees well with the simulated values, e.g. measured 60.2\,mV/fC and simulated 58.5\,mV/fC for RC=3. 
The example of the noise occupancy scan for the bare channel and RC set to 3 is shown in Figure \ref{fig:noiseOccupancy}. From this figure, one can extract not only the output noise but also the peaking time of the shaper response. 
The latter can be estimated from the maximum frequency of the noise hits at zero threshold using the Rice formula \cite{RiceFormula}. 
A summary of this theorem from the standpoint of the front end signal processing chains can be found in \cite{rivettiFEs}. 
The peaking time calculated for this particular setting of the RC filter is around 5.3\,ns which agrees well with the simulated value. The 2.9\,mV output noise combined with the channel gain of 60\,mV/fC gives the ENC of 300\,e$^-$ which has to be compared with the simulated 320\,e$^-$. Figure \ref{fig:twalk} shows the measurement of the time walk for the signals between 1.2 and 11\,fC for the threshold of 1\,fC. The 4.5\,ns time walk agrees well with the simulated numbers.

\section{Summary}
The FBCM23, a CMOS 65\,nm radiation tolerant readout ASIC has been designed and fabricated for the FBCM detector of the CMS experiment.   
The initial characterization of the ASIC confirms that the FBCM23 performance in terms of time resolution and noise levels is within the specifications.
A more elaborated readout system using CMS tracker data acquisition modules is under development. It will allow for testing simultaneously of all ASIC channels i.e., it will provide higher statistics in a shorter time and will be also used in the upcoming test beams. The X-ray irradiation of the FBCM23 ASIC is planned in the coming weeks.

\bibliographystyle{JHEP}
\bibliography{main.bib}

\providecommand{\href}[2]{#2}\begingroup\raggedright\begin{thebibliography}{10}

\bibitem{Przyborowski:2016aco}
D.~Przyborowski, J.~Kaplon and P.~Rymaszewski, \emph{{Design and Performance of
  the BCM1F Front End ASIC for the Beam Condition Monitoring System at the CMS
  Experiment}}, \href{https://doi.org/10.1109/TNS.2016.2575781}{\emph{IEEE
  Trans. Nucl. Sci.} {\bfseries 63} (2016) 2300}.

\bibitem{Sedghi_2022}
M.~Sedghi and on~behalf of~the CMS~Collaboration, \emph{Fast beam condition
  monitor of the cms experiment at hl-lhc},
  \href{https://doi.org/10.1088/1742-6596/2374/1/012011}{\emph{Journal of
  Physics: Conference Series} {\bfseries 2374} (2022) 012011}.

\bibitem{Collaboration:2759074}
CMS, \emph{{The Phase-2 Upgrade of the CMS Beam Radiation Instrumentation and
  Luminosity Detectors}},  Tech. Rep.
  \href{https://cds.cern.ch/record/2759074}{CERN-LHCC-2021-008, CMS-TDR-023},
  CERN, Geneva (2021).

\bibitem{LINDSTROM200330}
G.~Lindström, \emph{Radiation damage in silicon detectors},
  \href{https://doi.org/https://doi.org/10.1016/S0168-9002(03)01874-6}{\emph{Nucl.
  Instr. and Meth. A} {\bfseries 512} (2003) 30}.

\bibitem{WONSAK2015126}
S.~Wonsak et~al., \emph{Measurements of the reverse current of highly
  irradiated silicon sensors},
  \href{https://doi.org/https://doi.org/10.1016/j.nima.2015.04.027}{\emph{Nucl.
  Instr. and Meth. A} {\bfseries 796} (2015) 126}.

\bibitem{Akchurin_2020}
N.~Akchurin et~al., \emph{Charge collection and electrical characterization of
  neutron irradiated silicon pad detectors for the cms high granularity
  calorimeter},
  \href{https://doi.org/10.1088/1748-0221/15/09/P09031}{\emph{Journal of
  Instrumentation} {\bfseries 15} (2020) P09031}.

\bibitem{1589268}
J.~Kaplon and W.~Dabrowski, \emph{Fast cmos binary front end for silicon strip
  detectors at lhc experiments},
  \href{https://doi.org/10.1109/TNS.2005.862826}{\emph{IEEE Transactions on
  Nuclear Science} {\bfseries 52} (2005) 2713}.

\bibitem{6220879}
J.~Kaplon and M.~Noy, \emph{Front end electronics for slhc semiconductor
  trackers in cmos 90 nm and 130 nm processes},
  \href{https://doi.org/10.1109/TNS.2012.2200503}{\emph{IEEE Transactions on
  Nuclear Science} {\bfseries 59} (2012) 1611}.

\bibitem{Moreira:2809058}
P.~Moreira et~al., \emph{{lpGBT documentation: release}} (2022).

\bibitem{RiceFormula}
S.O.~Rice, \emph{Mathematical analysis of random noise},
  \href{https://doi.org/10.1002/j.1538-7305.1944.tb00874.x}{\emph{The Bell
  System Technical Journal} {\bfseries 23} (1944) 282}.

\bibitem{rivettiFEs}
A.~Rivetti, \emph{CMOS: Front-End Electronics for Radiation Sensors}, CRC Press
  (2015),
  \href{https://doi.org/https://doi.org/10.1201/b18599}{https://doi.org/10.1201/b18599}.

\end{thebibliography}\endgroup

\end{document}